\documentclass[preprint]{article}

\usepackage{jcappub}
\usepackage{hyperref}
\usepackage{physics}
\usepackage{ascmac}
\usepackage{color}
\usepackage{tcolorbox}
\usepackage{tikz}
\usetikzlibrary{intersections, calc, patterns, arrows.meta, positioning}

\newtheorem{definition}{Definition}
\newtheorem{conjecture}{Conjecture}

\begin{document}

\noindent KEK-TH-2777, KEK-Cosmo-0398

\arxivnumber{2511.15423} 
\title{A New Definition of Horndeski Theory and the Possibility of Multiple Scalar Field Extensions}


\author{Tomoki Katayama}
\affiliation{Institute of Particle and Nuclear Studies, High Energy Accelerator Research Organization (KEK), Oho 1-1, Tsukuba 305-0801, Japan}
\affiliation{The Graduate University for Advanced Studies (SOKENDAI), Tsukuba 305-0801, Japan }
\emailAdd{tomokika@post.kek.jp}


\abstract{In the single-field case, Horndeski provides the most general scalar–tensor theory with second-order field equations. By contrast, systematic multi-field extensions remain incomplete: while the general field equations for the bi-Horndeski case are known, a general action has not been established, and for cases with three or more fields, neither a general action nor general equations are available. We characterize Horndeski theory by three mild axioms: closure under invertible pure disformal transformations, the inclusion of a minimal Horndeski theory as an anchor, and the requirement that the coefficient functions of the Lagrangian be arbitrary up to relations required by the closure. Under this characterization, we recover the standard single-field action up to boundary terms and obtain a practical path to multi-field constructions. In particular, we show that antisymmetric structures, such as those identified by E. Allys, S.  Akama, and T. Kobayashi, appear within this framework, and indicate that this viewpoint has the potential to account for features captured by known bi-Horndeski equations.
}


\maketitle

\section{Introduction}

Scalar-tensor theory \cite{fujii2003scalar} is thought to be helpful in describing the classical dynamics of inflation \cite{Baumann:2009ds} and dark energy \cite{Copeland:2006wr,Ratra:1987rm,Wetterich:1987fm,Caldwell:1997ii,Zlatev:1998tr,Armendariz-Picon:2000ulo,Deffayet:2009wt,Deffayet:2009mn,DeFelice:2010pv,Deffayet:2010qz,Langlois:2018dxi}. It is known that Ostrogradsky ghosts \cite{Ostrogradsky:1850fid,Woodard:2015zca} generally appear when higher-order derivatives of scalar fields are included in the action of scalar-tensor theory \cite{Kobayashi:2019hrl,Lin:2017utd,Sivanesan:2013tba,Nicolis:2008in}, making construction difficult. The Horndeski theory \cite{Horndeski:1974wa} was introduced in 1974 as the most general single-field scalar-tensor theory whose equations of motion can be written in terms of second-order derivatives. It was rediscovered in 2011 \cite{Deffayet:2011gz,Kobayashi:2011nu}. Since then, theories beyond Horndeski, such as the GLPV theory \cite{Gleyzes:2013ooa,Gleyzes:2014dya}, the DHOST theory \cite{Langlois:2015cwa,BenAchour:2016cay,Crisostomi:2016czh,BenAchour:2016fzp}, and Theories that transcend them \cite{Takahashi:2022mew,Takahashi:2023jro,Michiwaki:2026xru}, have been discovered. These theories have been extensively studied \footnote{sometimes using the language of effective field theory \cite{Cheung:2007st,Gubitosi:2012hu,Gleyzes:2013ooa,Gleyzes:2014rba,Langlois:2017mxy,Frusciante:2019xia,Takahashi:2023jro}.}, including their verification in cosmology \cite{observations:2022mik,DeFelice:2010nf,Yamauchi:2017ibz,Langlois:2017dyl,Yamauchi:2021nxw,Yamauchi:2022fss,Sugiyama:2023tes,Yamashita:2024mdn}.

While the general theory of scalar–tensor theory for a single scalar field has achieved resounding success, the general theory of scalar–tensor theory for multiple scalar fields remains largely unexplored. Regarding the class extending Horndeski theory to multiple scalars, only the equations of motion for the case involving two scalar fields are known, while the action remains undiscovered \cite{Ohashi:2015fma}. Furthermore, for theories involving three or more scalar fields, both the equations of motion and the action remain undiscovered. However, there are previous studies, including those that did not succeed \cite{Horndeski:2024hee}. To date, several studies have identified terms \cite{Kobayashi:2013ina,Allys:2016hfl,Akama:2017jsa,Aoki:2021kla} that are never included when the scalar field is single. This paper refers to these terms as the Allys-Akama-Kobayashi terms. Therefore, a theory must at least be able to describe these terms to qualify as a multi-Horndeski theory.

Horndeski theory \cite{Horndeski:1974wa,Deffayet:2009mn,Kobayashi:2011nu} is a comprehensive scalar-tensor framework that incorporates a single scalar field and is characterized by equations of motion involving only second-order derivatives. Horndeski theory was formulated by extending Lovelock’s theorem \cite{Lovelock:1971yv,Lovelock:1972vz} to include a scalar field. Furthermore, as an extension to multiple fields, the equations of motion for bi-Horndeski theory (Horndeski theory with two scalar fields) are similarly constructed \cite{Ohashi:2015fma}, and research on constructing the action is currently underway \cite{Horndeski:2024hee,Nejati:2023hpe,Nejati:2024tuo}. However, deriving the multi-Horndeski theory using this method— the most general theory with an arbitrary number $\mathcal{N}$ of scalar fields, whose equations of motion can be written using up to second-order derivatives of each scalar field and the metric tensor —would be an extremely lengthy process.

Research extending scalar-tensor theory to multiple fields has entered a period of stagnation compared to about a decade ago. In addition, observations of dark energy have advanced significantly, with DESI results \cite{DESI:2024mwx,DESI:2025zgx} showing signs of inconsistency with the $\Lambda$CDM model \cite{Efstathiou:1990xe} that assumes the cosmological constant is dark energy. Extensions of scalar-tensor theory that incorporate multiple scalar fields to describe dynamic dark energy models from a broader range of possibilities are needed.

This paper provides a brief overview of single-field theory in Section 2 and multiple-field theory in Section 3. Section 4 introduces a new definition of the Horndeski theory and explains its construction under this definition. Section 5 demonstrates that the Allys-Akama-Kobayashi terms naturally emerge when extending a new definition of Horndeski theory to two fields.

\section{Single-Field Case}

\subsection{Horndeski theory}

In this section, let us consider Horndeski theory \cite{Horndeski:1974wa}. Note that for convenience, we will use the generalized Galileon theory \cite{Deffayet:2009mn,Deffayet:2011gz} for the action used here. Galileon theory \cite{Nicolis:2008in} is a theory of scalar fields invariant under transformation $\phi\to\phi+b_\mu x^\mu+c$ (where $b_\mu$ and $c$ are constants), and its equations of motion can be expressed using up to second-order derivatives. Generalized Galileon theory extends this to curved space-time \cite{Deffayet:2009wt} and generalizes the coefficients \cite{Deffayet:2009mn}. Furthermore, it has been shown that the generalized Galileon theory is equivalent to the Horndeski theory \cite{Kobayashi:2011nu}. Therefore, the action becomes 
\begin{align}
     S_H[\phi, g_{\mu\nu}]=S_{GG}[\phi, g_{\mu\nu}]=\int d^4x\sqrt{-g}\mathcal{L}_{GG}\, , \end{align}
 where $\mathcal{L}_{GG}$ is defined as follows, 
 \begin{align}
    \mathcal{L}_{GG}=&G_2(\phi, X)-G_3(\phi, X)\Box\phi+G_4(\phi, X)R+G_{4X}\bigl((\Box\phi)^2-\phi^\mu_\nu\phi^\nu_\mu\bigr) \notag \\
    &+G_5(\phi, X)G^{\mu\nu}\phi_{\mu\nu}-\frac{1}{6}G_{5X}\bigl((\Box\phi)^3-3\phi^\mu_\nu\phi^\nu_\mu\Box\phi+2\phi^\mu_\nu\phi^\nu_\rho\phi^\rho_\mu\bigr)\,, 
\end{align}
 where $G_2, G_3, G_4$ and $G_5$ are arbitrary functions of $\phi$ and $X: =-\phi^\mu\phi_\mu/2$ and where $\phi_{\mu}: =\nabla_\mu\phi, \phi_{\mu\nu}: =\nabla_\nu\nabla_\mu\phi$, $G_{iA}: =\partial G_i/\partial A$ and $R$ is the Ricci scalar, $G^{\mu\nu}$ is the Einstein tensor.

 \subsection{Disformal transformation}

In this subsection, we introduce invertible pure disformal transformations related to important properties of Horndeski theory. An invertible pure disformal transformation \footnote{Note that terminology about the detailed classification of disformal conversion varies across different papers. In this paper, we refer to those with neither factor showing $X$ dependence as ``pure", and those with both factors showing $X$ dependence as ``full". It is also sometimes referred to as a $\phi$-dependent disformal transformation.} is defined as a frame transformation
        \begin{align}
            &g_{\mu\nu}\to\tilde{g}_{\mu\nu}=A(\phi)g_{\mu\nu}+B(\phi)\phi_\mu\phi_\nu\, ,\\
            &A\neq0\, , A-2BX\neq0\, .
        \end{align}

Bekenstein originally introduced this transformation \cite{Bekenstein:1992pj}, which was of the form $g_{\mu\nu}\to\tilde{g}_{\mu\nu}=A(\phi, X)g_{\mu\nu}+B(\phi, X)\phi_\mu\phi_\nu$. We refer to $A$ as the conformal factor and $B$ as the disformal factor. The term ``pure" here refers to the case where these two factors do not depend on $X$. At the same time, ``invertible" indicates that we are only considering disformal transformations for which an invertible transformation exists. This transformation is very commonly used in the context of scalar–tensor theory, and is typically employed in the same manner as the well-known procedure for converting from non-minimal to minimal coupled systems via conformal transformations. 
While this transformation has yielded significant insight \cite{Bettoni:2013diz,Zumalacarregui:2013pma,Gleyzes:2014qga} into scalar–tensor theory, we will focus here solely on introducing key properties of Horndeski theory.

 Applying this transformation to the Horndeski theory yields a transformation of the form
 \begin{align}
 S_H[\phi, g_{\mu\nu}]\to S_H[\phi,\tilde{g}_{\mu\nu}]=S_H'[\phi, g_{\mu\nu}] \, .
 \end{align}
 It is known that applying an invertible pure disformal transformation to Horndeski theory results in a return to Horndeski theory except for the boundary terms \cite{Bettoni:2013diz}.

\section{Multi-Field Case}

\subsection{Generalized Multi-Galileon Theory}

Horndeski theory was introduced as the most general theory that includes one scalar field and a metric tensor field up to their second-order derivatives in the equations of motion. Here, we present the multi-Horndeski theory as the most general theory that includes $\mathcal{N}$ scalar fields and a metric tensor field up to their second-order derivatives in the equations of motion. In this subsection, we consider the most straightforward candidate for a multi-Horndeski theory, namely the ``Generalized multi-Galileon theory" \cite{Padilla:2012dx,Sivanesan:2013tba,Kobayashi:2013ina}, which is a multi-field version of the generalized Galileon theory, and its action is
\begin{align}
    S_{\rm{GMG}}[\phi^I, g_{\mu\nu}]&=\int d^4x\sqrt{-g}\mathcal{L}_{\rm{GMG}}\, , \\
    \mathcal{L}_{\rm{GMG}}&=\sum^5_{n=2}\mathcal{L}_{n}\, , 
\end{align}
where each $\mathcal{L}_i$ takes the form
 \begin{align}
 \mathcal{L}_2: =&G_2(\phi^A, X^{BC})\, , \\
 \mathcal{L}_3: =&-G_{3I}(\phi^A, X^{BC})\Box\phi^I\, ,\\
 \mathcal{L}_4: =&G_4(\phi^A, X^{BC})R+G_{4,\langle IJ\rangle}(\phi^A, X^{BC})\big(\Box\phi^I\Box\phi^J-\phi^{I\mu}_{\nu}\phi^{J\nu}_{\mu}\big)\, ,\\
 \mathcal{L}_5: =&G_{5I}(\phi^A, X^{BC})G_{\mu\nu}\phi^{I\mu\nu} \notag \\
 &-\frac{1}{6}G_{5I,\langle JK\rangle}(\phi^A, X^{BC})\big(\Box\phi^{(I}\Box\phi^J\Box\phi^{K)}-3\Box\phi^{(I}\phi^{J\mu}_{\nu}\phi^{K)\nu}_{\mu}+2\phi^{(I\mu}_{\lambda}\phi^{J\nu}_{\mu}\phi^{K)\lambda}_\nu\big)\, .
 \end{align}
 the capital indices $A, B,\cdots=1, 2,\cdots,\mathcal{N}$ label the scalar field $\phi^I$. Here, we are using the quantities defined as $X^{BC}: =-\phi^{(B\mu}\phi^{C)}_\mu/2$ and $F_{, \langle JK\rangle}: =(\partial F/\partial X^{JK}+\partial F/\partial X^{KJ})/2$ \footnote{This notation has been employed in prior research \cite{Langlois:2008qf}.}.

 At first glance, this may appear to express the multi-Horndeski theory, but it does not represent all terms.

\subsection{Extended multi-Galileon theory}

 A theory called the extended multi-Galileon theory \cite{Allys:2016hfl} was discussed to examine terms not included in the previously mentioned generalized multi-Galileon theory. For example, term
 \begin{align}
 \mathcal{L}_{\text{add}}\propto \delta_{I[J}\delta_{K]L}\delta^{\mu_1\mu_2\mu_3\mu_4}_{\nu_1\nu_2\nu_3\nu_4}\phi^{I\nu_1}\phi^{L\nu_2}\phi^K_{\mu_1}\phi^J_{\mu_2}R^{\nu_3\nu_4}{}_{\mu_3\mu_4} \, , \end{align}
  where $\delta^{\mu_1\cdots\mu_n}_{\nu_1\cdots\nu_n}$ is defined as $n!\delta^{[\mu_1}_{\nu_1}\cdots\delta^{\mu_n]}_{\nu_n}$,
  derived within the multi-DBI Galileon theory \cite{Renaux-Petel:2011rmu}, is not included in the generalized multi-Galileon theory \cite{Kobayashi:2013ina}, but it is known that its equations of motion can be written up to second-order derivatives.

 Furthermore, previous research has indicated that at least the Allys-Akama-Kobayashi terms (AAK terms) \footnote{Refer to Appendix C for details.} should be included to satisfy the multi-Horndeski theory \cite{Allys:2016hfl,Akama:2017jsa}. The AAK terms exhibit antisymmetry in their internal indices and do not appear trivially in Horndeski theory; they are terms specific to theories with multiple scalar fields. The AAK terms are of the form 
 \begin{align}
 \mathcal{L}_{\rm{AAK}1}=&A_{[IJ][KL]M}(\phi^A, X^{BC})\delta^{\mu_1\mu_2\mu_3}_{\nu_1\nu_2\nu_3}\phi^I_{\mu_1}\phi^J_{\mu_2}\phi^{K\nu_1}\phi^{L\nu_2}\phi^{M\nu_3}_{\mu_3}\, ,\\
 \mathcal{L}_{\rm{AAK}2}=&B_{[IJ][KL]}(\phi^A, X^{BC})\delta^{\mu_1\mu_2\mu_3\mu_4}_{\nu_1\nu_2\nu_3\nu_4}\phi^I_{\mu_1}\phi^J_{\mu_2}\phi^{K\nu_1}\phi^{L\nu_2}R^{\nu_3\nu_4}{}_{\mu_3\mu_4} \notag \\
 &+2B_{[IJ][KL],\langle MN\rangle}\delta^{\mu_1\mu_2\mu_3\mu_4}_{\nu_1\nu_2\nu_3\nu_4}\phi^I_{\mu_1}\phi^J_{\mu_2}\phi^{K\nu_1}\phi^{L\nu_2}\phi^{M\nu_3}_{\mu_3}\phi^{N\nu_4}_{\mu_4}\,, \\
 \mathcal{L}_{\rm{AAK}3}=&C_{[IJK][LMN]O}(\phi^A, X^{BC})\delta^{\mu_1\mu_2\mu_3\mu_4}_{\nu_1\nu_2\nu_3\nu_4}\phi^I_{\mu_1}\phi^J_{\mu_2}\phi^K_{\mu_3}\phi^{L\nu_1}\phi^{M\nu_2}\phi^{N\nu_3}\phi^{O\nu_4}_{\mu_4}\, , \end{align}

 Here, it is immediately apparent that $\mathcal{L}_{\rm{add}}$ actually contains $\mathcal{L}_{\rm{AAK}2}$ in the form of $B_{[IJ][KL]}\propto\delta_{I[K}\delta_{L]J}$. It is clear that the AAK terms incorporated into the generalized multi-Galileon theory are essential elements of the multi-Horndeski framework. Furthermore, as shown in prior research \cite{Ohashi:2015fma}, these terms fill in the missing basis in the equations of motion.

\subsection{Challenges in Constructing the Multi-Horndeski Theory}

We have a complete understanding of the equations of motion of the bi-Horndeski theory \cite{Ohashi:2015fma}. These equations were derived using a method that fully extends Horndeski's approach \cite{Horndeski:1974wa}. However, the action involving two scalar fields, as well as the actions and equations of motion involving three or more scalar fields, remain poorly understood. This is because deriving the action from the complex equations of motion is difficult, and even now, nine years after the successful derivation of the equations of motion, no paper attempting to prove it \cite{Horndeski:2024hee} has been published.

Furthermore, research on the equations of motion for $\mathcal{N}\geq3$ has not continued, as they become even more complex. As a result, this approach is very costly and not easy. On the other hand, extending the generalized Galileon theory, which is equivalent to Horndeski theory, to multiple scalar fields has also not worked well in practice. There is no guarantee that the multi-Horndeski theory and the generalized multi-Galileon theory are equivalent.

\section{New definition of the Horndeski theory}

\subsection{Introduction of a new definition}

 In this paper, we propose a new definition of Horndeski theory that does not rely on specific actions. It is given as follows.
 \begin{tcolorbox}[colframe=red, colback=red!3!]
 \begin{definition}
 Horndeski theory is defined as the minimal class of Lagrangians that satisfies the following

 \begin{enumerate}
 \item Closed under ``invertible pure disformal transformations".
 \item Contains ``minimal Horndeski theory".
 \item Each term of a Lagrangian in the class is a coefficient function of $\phi$ and $X$ times a tensor structure that contains the second or higher derivatives of $\phi$ or the curvature tensors, or is trivial. The coefficient functions are arbitrary — the class includes all Lagrangians with any choice of them — except that some coefficients may be fixed by those of other terms through the relations required by condition 1.
 \end{enumerate}
 \end{definition}
 \end{tcolorbox}

 \begin{tcolorbox}[colframe=red, colback=red!3!]
 \begin{definition}
 The minimal Horndeski theory is defined as
 \begin{align}
 \alpha X+R+\beta(\phi)G^{\mu\nu}\phi_{\mu\nu}\,,
 \end{align}
 where $\alpha$ is a constant, and $\beta(\phi)$ is an arbitrary function of $\phi$.

 \end{definition}
 \end{tcolorbox}

 Here, the action established by this definition is uniquely identified as $S_H$, aside from boundary terms. \footnote{Uniqueness here means uniqueness up to boundary terms. The conditions in the definition are understood at the level of Lagrangian representatives: Lagrangians differing by boundary terms are regarded as distinct representatives, and in particular the anchor (4.1) is specified in its un-integrated form (see Appendix A.2 for why this matters). Since at least one complete form of the Horndeski action is known, we take it as the standard representative.}. One method for deriving an action that satisfies this definition is described in Appendix A; please refer to it for additional details. Under the present definition, ``deriving all terms of the most general equation of motion" is not a definition of Horndeski theory but a property, and it is a non-trivial claim. However, since the same action \cite{Deffayet:2011gz} is actually derived, it can be confirmed that this ``property” holds.

 Regarding the second point in this definition, ``Contains $\alpha X+R+\beta(\phi)G^{\mu\nu}\phi_{\mu\nu}$.", we should clarify that this is not to claim that ``the Horndeski theory definition is obtained by mapping this Lagrangian via an invertible pure disformal transformation.". In DHOST theory, it is known that there exist multiple classes closed under invertible disformal transformations. Therefore, this statement serves as an ``anchor" specifying which class the Horndeski theory resides in. 

 Furthermore, the third condition allows the coefficients to be arbitrary without spoiling the absence of Ostrogradsky ghosts, since the relations among the coefficients are retained.

\subsection{Minimal Horndeski Theory}

First, we will address the reasons why $\alpha X+R+\beta(\phi)G^{\mu\nu}\phi_{\mu\nu}$ is classified as the minimal Horndeski theory in this context. Primarily, within the framework of gravitational theory, the fundamental aspect of scalar-tensor theory pertains to the behavior of the tensor. Therefore, to encapsulate general relativity, it must initially incorporate $R$. Nonetheless, the constant multiplier of $R$ is regarded as an overarching term and is consequently omitted. Moreover, in the context of general relativity, the scalar field is not required to be incorporated within the coefficient component of $R$. Next, to explicitly incorporate the dynamics of the scalar field, we introduce the momentum term for the scalar field. However, due to the lack of a fixed constant ratio between this factor and the coefficient of $R$, we introduce the parameter $\alpha$. The term $\alpha X$ is incorporated exclusively to explicitly depict the dynamics of the scalar field and produces solely terms associated with $G_2$ in the disformal transformation. Therefore, most generated terms originate from $R$ \cite{Bettoni:2013diz,Alinea:2020sei}. The third term involves a somewhat intricate discussion, details of which are provided in Appendix A.2; it is a technical term introduced to naturally generate terms related to $G_5$. For the reasons above, we argue that the parameterization in $\alpha$ and $\beta(\phi)$ represents the ``minimal” theory that reproduces Horndeski theory.

\subsection{Why the new definition?}



Let us clarify what we mean by a ``definition'' in this work. 
We do give a definition of Horndeski theory; however, it is not the conventional definition formulated in terms of ``most general equations of motion containing up to second derivatives.'' 
Instead, we adopt a different set of defining principles motivated by disformal geometry: we characterize the theory by closure under invertible pure disformal transformations, an anchor condition specifying the class, and the arbitrariness of the coefficient functions up to relations required by the closure; together these reproduce the standard single-field Horndeski/generalized Galileon action (up to boundary terms).

In this sense, our definition is not obtained by imposing equations of motion that include up to the most general second-order differential equations.
Rather, it provides an alternative (and constructive) definition of the same theory, designed to make the disformal structure manifest and to serve as a natural starting point for systematic multi-field generalizations.

This methodology diverges from the traditional approach of deriving the action directly from the equations of motion, as proposed by Horndeski \textit{et al.} \cite{Horndeski:1974wa,Ohashi:2015fma,Horndeski:2024hee} in the aforementioned context. It facilitates the derivation of the action without necessitating a discussion of the equations of motion and can be seamlessly extended to encompass the case of an arbitrary number of $\mathcal{N}$ scalar fields.

\section{Application of the New Definition of 
Horndeski Theory to Multiple Fields}

\subsection{A New Definition of the Multi-Horndeski Theory}

The advantage of handling Horndeski theory with the new definition is that it provides a straightforward way to extend the theory. For example, with the traditional definition, although it was easy to extend to multiple scalar fields, actually calculating the action was very difficult, effectively limiting the definition to the action itself. In contrast, the new definition only requires extending the ``disformal transformation", the ``minimal Horndeski theory", and the arguments of the coefficient functions in the third condition to multiple fields\footnote{\cite{Domenech:2025gao} also discusses disformal transformations involving two scalar fields.}, instead of the entire action. This approach simplifies the mathematics and makes the derivation easier.

\subsection{Conjecture on the New Definition of the multi-Horndeski Theory}

In extending the new definition of the Horndeski theory to multiple fields, we present the most important conjecture.

\begin{tcolorbox}[colframe=red, colback=red!3!]
 \begin{conjecture}
 The multi-Horndeski theory, under the new definition, yields the most general equations of motion, including up to second-order derivatives of multiple scalar fields and the metric tensor.
\end{conjecture}
\end{tcolorbox}

If this conjecture proves correct, it would enable the construction of actions without imposing the condition that Horndeski theory possesses ``the most general equations including second-order derivatives”. Naturally, this hypothesis is not trivial, and its validity is not a priori guaranteed. Nevertheless, it holds for the $\mathcal{N}=1$ case, Horndeski theory. Furthermore, for the $\mathcal{N}=2$ case (i.e., the bi-Horndeski theory), since the equations of motion have already been derived, we need only verify whether we can derive equations equivalent to them. However, we defer to future work the construction of the complete bi-Horndeski action including the cubic sector. In the present paper, we confirm only that the quadratic truncation contains the AAK terms, and that the resulting equations are consistent with the sector of the known bi-Horndeski equations in which $K_I=0$.

However, it should be noted that, unlike in the case of a single scalar field, the closure property under multiple disformal transformations in multidimensional Horndeski theory has not yet been established. In this paper, we have adopted a construction that requires this property.

\subsection{Allys-Akama-Kobayashi terms from the New Definition}
 In this subsection, we confirm that the Allys-Akama-Kobayashi terms and the non-trivial effect of multiple scalar fields naturally appear in the multiple-scalar-field extension of Horndeski theory defined here.

 Next, we discuss the possibility of easily extending the new definition of the Horndeski theory to multiple fields as a concrete application. For simplicity, we consider the case of two scalar fields for which the equations of motion have already been obtained \cite{Ohashi:2015fma}. We define the bi-Horndeski theory in the same way as the Horndeski theory \cite{Katayama:2026aaa}. When extending the single-field definition adopted here to multiple fields, the elements requiring extension are the disformal transformation, the minimal Horndeski theory, and the arguments of the coefficient functions in the third condition. For the former, we define it as
 \begin{align}
 g_{\mu\nu}\to\tilde{g}_{\mu\nu}=A(\phi^A)g_{\mu\nu}+B_{IJ}(\phi^A)\phi^I_\mu\phi^J_\nu\, , \end{align}
based on previous research \cite{Watanabe:2015uqa,Firouzjahi:2018xob}\footnote{Here, the condition for an invertible transformation is $A\neq0,\,\det{(A\delta^I_J-2B_{JK}X^{IK})}\neq0$.}, and for the latter, we define it very simply as
 \begin{align}
 \alpha_{IJ}X^{IJ}+R+\beta_I(\phi^A)G^{\mu\nu}\phi^I_{\mu\nu}\, ,
 \end{align}
 where $\alpha_{IJ}$ is a constant, and $\beta_I(\phi^A)$ is an arbitrary function of $\phi^A$.
 Such an extension allows us to define the Horndeski theory containing two scalar fields.

 \begin{tcolorbox}[colframe=red, colback=red!3!]
 \begin{definition}
 The bi-Horndeski theory is defined as the minimal class of Lagrangians that satisfies the following

 \begin{enumerate}
 \item Closed under ``invertible pure bi-disformal transformations".
 \item Contains the ``minimal bi-Horndeski theory".
 \item Each term of a Lagrangian in the class is a coefficient function of $\phi^A$ and $X^{BC}$ times a tensor structure that contains the second or higher derivatives of $\phi^A$ or the curvature tensors, or is trivial. The coefficient functions are arbitrary functions with fixed index symmetries — the class includes all Lagrangians with any choice of them — except that some coefficients may be fixed by those of other terms through the relations required by condition 1.
 \end{enumerate}
 \end{definition}
 \end{tcolorbox}

In this work, since we are specifically considering up to $\mathcal{L}_4$, we set $\beta_I=0$ and $\mathcal{L}_5$ is not being treated, because this simplification makes the calculation tractable while preserving the essential structure. We discuss the derivation of a complete bi-Horndeski theory under this new definition, and will address its equivalence to the actual bi-Horndeski theory in future work \cite{Katayama:2026aaa}.

The action of the bi-Horndeski theory up to $\mathcal{L}_4$ derived under this definition is
\begin{align}
 \mathcal{L}_{\rm{qbH}}=&\sum^4_{n=2}\mathcal{L}_{n}+\mathcal{L}_{\rm{extra}}\, ,
\end{align}
\begin{align}
 \mathcal{L}_{{\rm{extra}}}=&H_{1[IJ][KL]M}(\phi^A, X^{BC})\delta^{\mu_1\mu_2\mu_3}_{\nu_1\nu_2\nu_3}\phi^I_{\mu_1}\phi^J_{\mu_2}\phi^{K\nu_1}\phi^{L\nu_2}\phi^{M\nu_3}_{\mu_3}\\
 &+H_{2[IJ][KL]}(\phi^A, X^{BC})\delta^{\alpha_1\alpha_2\alpha_3\alpha_4}_{\beta_1\beta_2\beta_3\beta_4}\phi^I_{\alpha_1}\phi^J_{\alpha_2}\phi^{K\beta_1}\phi^{L\beta_2}R_{\alpha_3\alpha_4}{}^{\beta_3\beta_4} \notag \\
 &+2H_{2[IJ][KL], \langle MN\rangle}\delta^{\alpha_1\alpha_2\alpha_3\alpha_4}_{\beta_1\beta_2\beta_3\beta_4}\phi^I_{\alpha_1}\phi^J_{\alpha_2}\phi^{K\beta_1}\phi^{L\beta_2}\phi^{M\beta_3}_{\alpha_3}\phi^{N\beta_4}_{\alpha_4}\,.
\end{align}
 Here, it is immediately apparent that the term $\mathcal{L}_{\rm{extra}}$ is equal to $\mathcal{L}_{\rm{AAK}1}+\mathcal{L}_{\rm{AAK}2}$\footnote{For results up to $\mathcal{L}_4$, \cite{Alinea:2020sei} can serve as a reference for the calculation process.}. As for $\mathcal{L}_{AAK3}$, it only appears under condition $\mathcal{N}\geq3$ and is not present this time.

\subsection{Quadratic bi-Horndeski theory}

The quadratic bi-Horndeski theory under this definition is given by 
\begin{align}
    \mathcal{L}_{\rm{qbH}}=&G_2-G_{3I}\Box\phi^I+G_4R+G_{4,\langle IJ\rangle}(\Box\phi^I\Box\phi^J-\phi^{I\mu}_\nu\phi^{J\nu}_\mu) \notag \\
    &+H_{1[IJ][KL]M}(\phi^A, X^{BC})\delta^{\mu_1\mu_2\mu_3}_{\nu_1\nu_2\nu_3}\phi^I_{\mu_1}\phi^J_{\mu_2}\phi^{K\nu_1}\phi^{L\nu_2}\phi^{M\nu_3}_{\mu_3}\\
    &+H_{2[IJ][KL]}\delta^{\alpha_1\alpha_2\alpha_3\alpha_4}_{\beta_1\beta_2\beta_3\beta_4}\phi^I_{\alpha_1}\phi^J_{\alpha_2}\phi^{K\beta_1}\phi^{L\beta_2}R_{\alpha_3\alpha_4}{}^{\beta_3\beta_4} \notag \\
 &+2H_{2[IJ][KL], \langle MN\rangle}\delta^{\alpha_1\alpha_2\alpha_3\alpha_4}_{\beta_1\beta_2\beta_3\beta_4}\phi^I_{\alpha_1}\phi^J_{\alpha_2}\phi^{K\beta_1}\phi^{L\beta_2}\phi^{M\beta_3}_{\alpha_3}\phi^{N\beta_4}_{\alpha_4}\,.
\end{align}

Details of the equations of motion for the quadratic bi-Horndeski theory are presented in Appendix E. At the stage of the quadratic bi-Horndeski theory, this corresponds to the case where $E_{IJKLM} \neq 0$, which was missing in the generalized multi-Galileon theory discussed in the previous study \cite{Ohashi:2015fma}. Although $K_I = 0$, this term does not appear in this work because it originates from the cubic terms of the generalized multi-Galileon theory. However, since $K_I$ originates from the cubic sector, the structure of the present framework suggests that terms beyond the AAK terms (for example, the cubic AAK terms\footnote{The AAK terms are included in the quadratic bi-Horndeski theory and are naturally derived from the definition used in this paper. Similarly, the terms specific to multiple scalar fields contained in the cubic bi-Horndeski theory, which also follow from this definition, are referred to as the cubic AAK terms.}) naturally arise at cubic order.

This paper does not address the correspondence between individual terms and previous research \cite{Horndeski:2024hee}. The reason is that the scope of this paper is to verify consistency with the equations of motion, and this correspondence lies outside that scope. Investigation of this correspondence should be undertaken after the derivation of the full cubic action. \footnote{Future papers \cite{Katayama:2026aaa} will discuss the action of the full cubic bi-Horndeski.}

\section{Conclusion}
This research shows that extending Horndeski theory to multi-Horndeski theory is difficult, given its conventional definition and the construction of actions derived from it. It introduces a new definition of Horndeski theory explicitly designed for extension to multiple fields. This definition is constructed to yield the same action as the conventional Horndeski theory. Furthermore, under this definition, the multi-Horndeski theory, extended to multiple fields, naturally incorporates the Allys-Akama-Kobayashi terms. Therefore, using this definition to construct the multi-Horndeski theory is expected to be a beneficial approach. 

\section*{Acknowledgments}

Shun Arai and Atsushi Naruko contributed valuable discussions from the initial conception of this study. Hiromasa Tajima and Riku Yoshimoto also kindly assisted with lengthy and tedious calculations. Kazufumi Takahashi provided extremely valuable comments regarding this research. Daisuke Yamauchi provided invaluable assistance down to the finest details in the preparation of this paper. Furthermore, during the concluding phases of manuscript preparation, Kazuya Koyama reviewed the draft. During the revision process, I received extensive comments from David Langlois. This paper was supported in part by The Graduate University for Advanced Studies, SOKENDAI (SDP233302, SDP252301, and SDP253206). The author thanks the ``Atom-type Researcher" program at the Yukawa Institute for Theoretical Physics, Kyoto University, which provided the initial inspiration for this research.


%

\appendix

\section{Concrete method to construct actions from definitions}


This section introduces a constructive generating algorithm for the class of actions conforming to the definition provided in Sec 4.
Specifically, the algorithm accepts as an 
input an action $S_n$ within the class and outputs a new action $S_{n+1}$ belonging to the same class, thereby establishing an iterative sequence $S_n$ starting from an initial seed $S_1$. The methodology is founded on invertible (pure) disformal transformations and a structure-preserving generalization of the coefficient functions. Given that disformal manipulations can be algebraically extensive, we do not reproduce every intermediate identity. When standard disformal formulas are employed, the reader is referred to Appendix~A of Ref.~\cite{Bettoni:2013diz}.

\subsection{Generalized Disformal Mapping Method}

Our update rule, denoted as $S_n \mapsto S_{n+1}$, is referred to as the \emph{generalized disformal mapping method}. This approach involves two principal operations.

First, an \textbf{invertible disformal mapping} is performed. Starting from $S_n$, we apply an invertible \emph{pure} disformal transformation and obtain a new class of actions, which we call the \emph{disformal image} of $S_n$. Within this image, the dependence on $\phi$ and $X$ can be parameterized by a finite set of coefficient functions $f_i(\phi,X)$.

Second, we carry out a \textbf{structure-preserving generalization}. The disformal mapping does not produce arbitrary independent functions; instead, the coefficient functions $f_i(\phi,X)$ satisfy characteristic relations inherited from the transformation. When schematic relations of the form
\begin{equation}
  f_i(\phi,X) = a\,\partial_X f_j(\phi,X),
  \qquad a=\text{const.},
\end{equation}
are observed, we enlarge the class by promoting $f_i \to F_i$ while preserving the same differential relations, i.e.,
\begin{equation}
  F_i(\phi,X) = a\,\partial_X F_j(\phi,X).
\end{equation}

The resulting enlarged class is defined as $S_{n+1}$, which we refer to as the \emph{generalized disformal mapped} of $S_n$. This procedure mirrors the extension from the covariant Galileon \cite{Deffayet:2009wt} to the generalized Galileon \cite{Deffayet:2009mn}, wherein coefficient functions are generalized without modifying the derivative structure.

An essential aspect is that the coefficients are not fully generalized independently; rather, they are extended in a manner that preserves the differential relations mandated by the definition. The term “preserving differential relations” herein corresponds to “preserving the coefficient degeneracy condition” in DHOST. Consequently, it is anticipated that Ostrogradsky ghosts may be similarly eliminated even in the case of multiple scalar fields \footnote{However, no mathematical proof is furnished.}.

 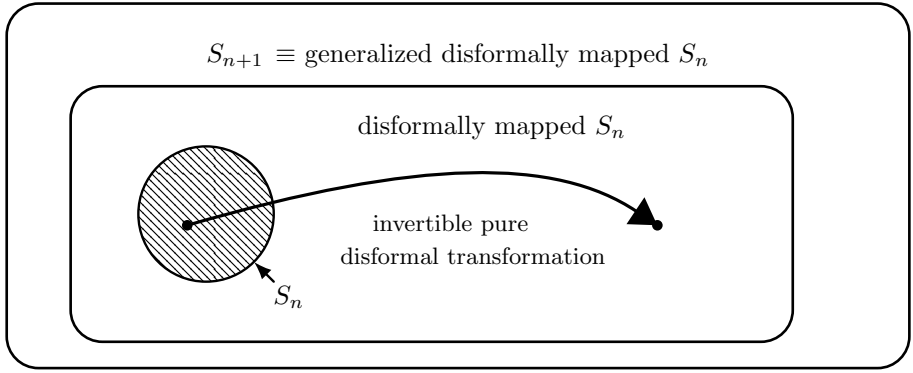
\begin{figure}[htbp]
 \centering
 \begin{tikzpicture}[line cap=round, line join=round, >=Triangle]
 \draw[line width=1pt, rounded corners=12pt]
 (-6.0, -4.0) rectangle (6.0, 0.85); \draw[line width=1pt, rounded corners=12pt]
 (-5.15, -3.65) rectangle (4.45, -0.25); \node at (0.0, 0.15) {$S_{n+1}\,\equiv$ generalized disformal mapped $S_n$}; \node at (0.45, -0.8) {disformal mapped $S_n$}; \filldraw[pattern=north west lines, pattern color=black, line width=0.9pt]
 (-3.35, -1.95) circle (0.90); \fill (-3.60, -2.10) circle (2pt); \node at (-2.25, -3.05) {$S_n$}; \draw[-{Latex[length=2.4mm]}, line width=0.9pt]
 (-2.45, -2.85) -- (-2.714, -2.586); \draw[-{Triangle[length=3.5mm, width=4.2mm]}, line width=1.2pt]
 (-3.60, -2.10).. controls (-0.50, -1.20) and (1.30, -1.20)..
 (2.65, -2.10); \fill (2.65, -2.10) circle (2pt); \node at (-0.1, -2.1) {\small invertible pure}; \node at (0.2, -2.5) {\small disformal transformation}; \end{tikzpicture}
 \caption{Conceptual figure to construct $S_{n+1}$ from $S_n$}
 \end{figure}

Next, we set $S_1$ to at least the actions included in the definition, in this case the minimal Horndeski theory $\alpha X+R+\beta(\phi)G_{\mu\nu}\phi^{\mu\nu}$.

Finally, if condition $S_{n+1}=S_{n}$—that is, the state where no further extension of the action is possible—can be constructed, then this is the Horndeski theory we sought.

\subsection{Why introduce the minimal Horndeski?}

 For example, we will examine the construction of the Horndeski theory when the general theory of relativity (the simplest scalar-tensor theory incorporating free scalar fields) is set on $S_1$.

 When we actually construct $S_2$, we can see that it indeed takes the same form as the Horndeski theory.
 \begin{align}
     S_2=\int d^4x\sqrt{-g}\Big(&f_1(\phi, X)-f_2(\phi, X)\Box\phi+f_3(\phi, X)R+f_{3X}\bigl((\Box\phi)^2-\phi^\mu_\nu\phi^\nu_\mu\bigr)\Big)\,.
 \end{align}
 Applying $S_3$ here yields $S_3=S_2$. In other words, the iteration ends at this stage. However, since this fixed point does not contain the $G_5$ sector, it reproduces only the quadratic Horndeski theory and cannot reproduce the full Horndeski theory. On the other hand, applying the well-known partial integration to $S_2$ yields 
 \begin{align}
     S_2'=\int d^4x\sqrt{-g}\Big(&f_1(\phi, X)-f_2(\phi, X)\Box\phi+(f_3(\phi, X)-X)R\notag \\
     &+(f_{3X}-1)\bigl((\Box\phi)^2-\phi^\mu_\nu\phi^\nu_\mu\bigr)-\phi G_{\mu\nu}\phi^{\mu\nu}\Big)\,.
 \end{align}
 By calculating $S_3$ in response to this, a different $S_3$ is obtained compared to before.
 \begin{align}
     S_3'=\int\sqrt{-g}d^4x\Big(&f_1(\phi, X)-f_2(\phi, X)\Box\phi+f_3(\phi, X)R+f_{3X}\bigl((\Box\phi)^2-\phi^\mu_\nu\phi^\nu_\mu\bigr) \notag \\
 &+f_4(\phi, X)G^{\mu\nu}\phi_{\mu\nu}-\frac{1}{6}f_{4X}\bigl((\Box\phi)^3-3\phi^\mu_\nu\phi^\nu_\mu\Box\phi+2\phi^\mu_\nu\phi^\nu_\rho\phi^\rho_\mu\bigr)\Big)\,.
 \end{align}
 From this, calculating $S_4$ yields $S_4=S_3'$, which is derived as the Horndeski theory.

As mentioned above, constructing the action concretely from $\mathcal{L}=\alpha X+R$ is somewhat cumbersome and requires preliminary transformation knowledge. Therefore, introducing terms already contained in $\mathcal{L}_5$ is convenient for us, who already know the complete action of Horndeski theory. Thus, we add $\beta(\phi)G_{\mu\nu}\phi^{\mu\nu}$ as the minimal additional term. By constructing it this way, executing the above process satisfies $S_3=S_2$, allowing us to generate the Horndeski theory action simply and mechanically. However, note that if $\beta$ is treated as a constant, this term becomes a boundary term.

\section{What is the meaning of the second condition in the definition?}

Here, we explain the importance of the second condition in the new definition, based on a well-known fact from DHOST theory \cite{Langlois:2017mdk,Langlois:2015cwa,BenAchour:2016fzp}. In DHOST theory, there are some classes based on the chosen degeneracy conditions \cite{BenAchour:2016fzp}, and Horndeski theory is part of the class known as disformal Horndeski theory. Each class remains invariant under an invertible full disformal transformation $g_{\mu\nu}\to\tilde{g}_{\mu\nu}=A(\phi,X)g_{\mu\nu}+B(\phi,X)\phi_\mu\phi_\nu$ . Therefore, it is believed that several
classes of theories are closed under invertible pure disformal transformations. Therefore, if only the first and third conditions of the definition are considered, each can be described as the Horndeski theory. Therefore, the second condition of the definition acts as an ``anchor” for determining which class constitutes the Horndeski theory.

\section{Allys-Akama-Kobayashi terms}

``Allys-Akama-Kobayashi terms" are terms discovered by Erwan Allys in the flat Galileon \cite{Allys:2016hfl} and extended to a covariant form by Shingo Akama and Tsutomu Kobayashi \cite{Akama:2017jsa}, possessing antisymmetry in their internal indices. They are therefore non-trivial terms that cannot arise from a simple multiple scalar field extension of Horndeski theory \cite{Kobayashi:2013ina}.

These terms contain the following three types of terms
\begin{align}
 \mathcal{L}_{\rm{AAK}1}=&A_{[IJ][KL]M}\delta^{\mu_1\mu_2\mu_3}_{\nu_1\nu_2\nu_3}\phi^I_{\mu_1}\phi^J_{\mu_2}\phi^{K\nu_1}\phi^{L\nu_2}\phi^{M\nu_3}_{\mu_3}\,,\\
 \mathcal{L}_{\rm{AAK}2}=&B_{[IJ][KL]}\delta^{\mu_1\mu_2\mu_3\mu_4}_{\nu_1\nu_2\nu_3\nu_4}\phi^I_{\mu_1}\phi^J_{\mu_2}\phi^{K\nu_1}\phi^{L\nu_2}R^{\nu_3\nu_4}_{\mu_3\mu_4}\notag\\
 &+2B_{[IJ][KL],\langle MN\rangle}\delta^{\mu_1\mu_2\mu_3\mu_4}_{\nu_1\nu_2\nu_3\nu_4}\phi^I_{\mu_1}\phi^J_{\mu_2}\phi^{K\nu_1}\phi^{L\nu_2}\phi^{M\nu_3}_{\mu_3}\phi^{N\nu_4}_{\mu_4}\,,\\
 \mathcal{L}_{\rm{AAK}3}=&C_{[IJK][LMN]O}\delta^{\mu_1\mu_2\mu_3\mu_4}_{\nu_1\nu_2\nu_3\nu_4}\phi^I_{\mu_1}\phi^J_{\mu_2}\phi^K_{\mu_3}\phi^{L\nu_1}\phi^{M\nu_2}\phi^{N\nu_3}\phi^{O\nu_4}_{\mu_4}\,.
\end{align}

\section{How do the Allys-Akama-Kobayashi terms appear?}

\subsection{AAK1}

First, we confirm that $\mathcal{L}_{AAK1}$ and $\mathcal{L}_{AAK3}$ consist of parts that can be included in $\mathcal{L}_2+\mathcal{L}_3$ and parts that cannot.\footnote{Since this discussion concerns the theory of two scalar fields, $\mathcal{L}_{AAK3}=0$; however, we are explicitly stating this because the same argument holds for three or more scalar fields.} If these terms are expanded using the generalized Kronecker delta, they become 
\begin{align}
 \mathcal{L}_{\rm{AAK}1}+\mathcal{L}_{\rm{AAK}3}=\alpha(\phi^A, X^{BC})+\beta_{I}(\phi^A, X^{BC})\Box\phi^I+\gamma_{IJK}(\phi^A, X^{BC})\phi^I_{\mu\nu}\phi^{J\mu}\phi^{K\nu}\, .
\end{align}
Recalling relation $\Theta_{, \langle IJ\rangle}\phi^{I\nu}\phi^J_{\nu\mu}=\Theta_{, I}\phi^I_\mu-\Theta(\phi^A, X^{BC})_{;\mu}$ where $\Theta$ is the arbitrary function of $\phi^A$ and $X^{BC}$ \footnote{Here, $\Theta_{,I}:=\partial\Theta/\partial\phi^I$.}. Here, we decompose $\gamma_{IJK}$ into part $\gamma_{\textrm{Int}IJK}$, which can be integrated with respect to $X^{IJ}$, and part $\gamma_{\textrm{non-Int}IJK}$, which cannot. If $\delta_I$ is defined as $\delta_I=\int dX^{JK}\gamma_{\textrm{Int}JKI}$, it can be rewritten as
\begin{align}
 \mathcal{L}_{\rm{AAK}1}+\mathcal{L}_{\rm{AAK}3}=&\alpha+\beta_{I}\Box\phi^I+\gamma_{\textrm{non-Int}IJK}(\phi^A, X^{BC})\phi^I_{\mu\nu}\phi^{J\mu}\phi^{K\nu}-2\delta_{I, J}X^{IJ}-\delta_{I;\mu}\phi^{I\mu}\, ,\notag\\
 =&\alpha'+\beta_{I}\Box\phi^I+\gamma_{\textrm{non-Int}IJK}(\phi^A, X^{BC})\phi^I_{\mu\nu}\phi^{J\mu}\phi^{K\nu}-\delta_{I;\mu}\phi^{I\mu}\, .
\end{align}
The last term yields $\delta_I\Box\phi^I$ by performing partial integration in the form of the action. Therefore, the AAK1 and AAK3 terms take the form of 
\begin{align}
    \mathcal{L}_{\rm{AAK}1}+\mathcal{L}_{\rm{AAK}3}=\alpha'(\phi^A,X^{BC})+\beta_{I}'(\phi^A,X^{BC})\Box\phi^I+\gamma_{\textrm{non-Int}IJK}(\phi^A, X^{BC})\phi^I_{\mu\nu}\phi^{J\mu}\phi^{K\nu}\,.
\end{align}
Therefore, in case $\gamma_{\textrm{non-Int}IJK}=0$, we can see that $\mathcal{L}_{\rm{AAK}1}+\mathcal{L}_{\rm{AAK}3}$ is completely contained within $\mathcal{L}_{2}+\mathcal{L}_{3}$.
Although the remaining terms cannot be immediately eliminated based on the index relations in the case of a single scalar field, since $\gamma_{\textrm{non-Int}IJK}=0$ is always equal to one in a single scalar field, terms of this form did not appear in Horndeski theory. Therefore, in the Multi-Horndeski theory, it generally appears independently as the AAK1 term.






\subsection{AAK2}

This appendix elucidates two key points. Firstly, it is demonstrated that the antisymmetry of indices emerges naturally. Secondly, it is shown that $\mathcal{L}_{\rm{AAK2}}$ is constructed utilizing the method outlined in Appendix A. To substantiate these assertions, a bi-disformal transformation is applied to the Einstein-Hilbert action, and the minimal necessary terms are computed for clarification. 
Additionally, it is noteworthy that more comprehensive calculations for the case of a single scalar field have been explicitly documented in prior research \cite{Bettoni:2013diz,Zumalacarregui:2013pma,Alinea:2020sei}, and these references should suffice.

After transforming the Christoffel $\tilde{\Gamma}^\mu_{\alpha\beta}$, focusing on the term that contains the second derivative of the scalar field, we find that it is 
\begin{align}
    \tilde{\Gamma}^\mu_{\alpha\beta}=\Gamma^\mu_{\alpha\beta}-\Big(\log{\sqrt{\mathcal{D}}}\Big){}_{,\langle IJ\rangle}\phi^{I\mu}\phi^J_{\alpha\beta}+\cdots\,,
\end{align}
where $\mathcal{D}:=1-2B_{IJ}X^{IJ}/A+4B_{I[J}B_{K]L}X^{IJ}X^{KL}/A^2$. For simplicity, we will refer to it as $D_{IJ}:=-(\log{\sqrt{\mathcal{D}}}){}_{,\langle IJ\rangle}$. Under this basis, the Riemann tensor includes third-order derivatives of the scalar field as follows.
\begin{align}
    \tilde{R}^\mu_{\nu\alpha\beta}=&R^\mu_{\nu\alpha\beta}-D_{IJ}\phi^{I\mu}\phi^J_{\nu[\alpha\beta]}+(D_{IJ,\langle KL\rangle}+D_{IJ}D_{KL})\phi^{I\mu}\phi^J_{\lambda[\alpha}\phi^{K\lambda}\phi^L_{\beta]\nu}\cdots\,,\notag \\
    =&R^\mu_{\nu\alpha\beta}-D_{IJ}\phi^{I\mu}\phi^{J\gamma}R_{\gamma\nu[\alpha\beta]}+(D_{IJ,\langle KL\rangle}+D_{IJ}D_{KL})\phi^{I\mu}\phi^J_{\lambda[\alpha}\phi^{K\lambda}\phi^L_{\beta]\nu}+\cdots\,.
\end{align}
From this point onward, it is clear that calculating the Ricci scalar yields an antisymmetric term. Furthermore, in light of symmetry and $D_{IJ,KL}=2D_{IK}D_{JL}$,
\begin{align}
    A\sqrt{\frac{\tilde{g}}{g}}\tilde{R}=&-\sqrt{\mathcal{D}}D_{IJ}D_{KL}\phi^{I\alpha}\phi^{J\beta}\phi^{K\gamma}\phi^{L\delta}R_{\beta\gamma\alpha\delta}\notag \\
    &+2\sqrt{\mathcal{D}}D_{IJ}D_{KL}D_{MN}\phi^{I\alpha}\phi^{J\beta}\phi^{K\gamma}\phi^{L\delta}\phi^M_{\beta[\alpha}\phi^N_{\delta]\gamma}+\cdots\,.
\end{align}
Firstly, it is evident that the antisymmetry of the indices is inherently derived from this expression. Additionally, it is immediately clear that the binomials presented herein correspond to the coefficients of the curvature and non-curvature components of $\mathcal{L}_{\rm{AAK2}}$ except for constant-factor multipliers, respectively (if this remains ambiguous, one may explicitly expand $\delta^{\mu_1\mu_2\mu_3\mu_4}_{\nu_1\nu_2\nu_3\nu_4}$). Furthermore, because these coefficients are interconnected by 
\begin{align}
    (-\sqrt{\mathcal{D}}D_{IJ}D_{KL}/A)_{,\langle MN\rangle}=\sqrt{\mathcal{D}}D_{IJ}D_{KL}D_{MN}/A\,,
\end{align}
as can be verified through calculation, the general expression for $\mathcal{L}_{\rm{AAK2}}$ is consequently obtained.

\section{Equations of motion for the quadratic bi-Horndeski theory}

In the equation of motion $\mathcal{E}^{\rm quad}_{\mu\nu}=0$, the tensor $\mathcal{E}^{\rm quad}_{\mu\nu}$ is 
\begin{flalign}
&\mathcal E^{\rm quad}_{\mu\nu}
=
\left(
-\frac12G_2
+G_{3(I,J)}X^{IJ}
-2G_{4,I,J}X^{IJ}
-8H_{1IJKLM,N}X^{IJ}X^{KL}X^{MN}
\right)g_{\mu\nu}
&&\nonumber\\
&\quad
+\Biggl[
-\frac12G_{2,IJ}
+G_{3(I,J)}
-G_{4,I,J}
+32H_{2KM[L|N|,(I],J)}X^{KL}X^{MN}
&&\nonumber\\
&\qquad
-8H_{1KLMNO,P,\langle IJ\rangle}X^{KL}X^{MN}X^{OP}
&&\nonumber\\
&\qquad
-8H_{1KLMNO,P}
\left(
\delta^K{}_{(I}\delta^L{}_{J)}X^{MN}X^{OP}
+X^{KL}\delta^M{}_{(I}\delta^N{}_{J)}X^{OP}
\right.
&&\nonumber\\
&\hspace{52mm}\left.
+X^{KL}X^{MN}\delta^O{}_{(I}\delta^P{}_{J)}
\right)
&&\nonumber\\
&\qquad
-24H_{1IK(JMN),L}X^{KL}X^{MN}
-16H_{1IKOMN,L,\langle PJ\rangle}X^{KL}X^{MN}X^{OP}
&&\nonumber\\
&\qquad
-2
\left(
\delta^{PQ}_{KM}\delta^{RS}_{IN}
+\delta^{PQ}_{IN}\delta^{RS}_{KM}
\right)
\left(
H_{1PRQSO,L,\langle TJ\rangle}
-2H_{1PROSQ,L,\langle TJ\rangle}
\right.
&&\nonumber\\
&\hspace{60mm}\left.
-2H_{1POQST,L,\langle RJ\rangle}
\right)
X^{OT}X^{KL}X^{MN}
&&\nonumber\\
&\qquad
-\left(
\delta^{PQ}_{IL}\delta^{RS}_{KM}
+\delta^{PQ}_{KM}\delta^{RS}_{IL}
\right)
\left(
H_{1PRQSN,J}
+H_{1PRNSQ,J}
+H_{1PRQNS,J}
\right)
X^{KL}X^{MN}
&&\nonumber\\
&\qquad
-\left(
\delta^{PQ}_{JL}\delta^{RS}_{KM}
+\delta^{PQ}_{KM}\delta^{RS}_{JL}
\right)
\left(
H_{1PRQSN,I}
+H_{1PRNSQ,I}
+H_{1PRQNS,I}
\right)
X^{KL}X^{MN}
\Biggr]\phi^I_\mu\phi^J_\nu
&&\nonumber\\
&\quad
+\Biggl[
-X^{JK}G_{3IJK}
+G_{4,I}
+2X^{JK}G_{4IJ,K}
+16H_{2KM[L|N|,I]}X^{KL}X^{MN}
&&\nonumber\\
&\qquad
-24H_{1JK(ILM)}X^{JK}X^{LM}
-16H_{1JKNLM,\langle OI\rangle}X^{JK}X^{LM}X^{NO}
&&\nonumber\\
&\qquad
-\left(
\delta^{PQ}_{JL}\delta^{RS}_{KM}
+\delta^{PQ}_{KM}\delta^{RS}_{JL}
\right)
\left(
H_{1PRQSN,\langle OI\rangle}
-2H_{1PRNSQ,\langle OI\rangle}
\right.
&&\nonumber\\
&\hspace{60mm}\left.
-2H_{1PNQSO,\langle RI\rangle}
\right)
X^{NO}X^{JK}X^{LM}
&&\nonumber\\
&\qquad
+\left(
\delta^{PQ}_{JL}\delta^{RS}_{KM}
+\delta^{PQ}_{KM}\delta^{RS}_{JL}
\right)
\left(
H_{1PRQSI}
+H_{1PRISQ}
+H_{1PRQIS}
\right)
X^{JK}X^{LM}
\Biggr]
g_{\lambda(\mu}
\delta^{\lambda\alpha}_{\nu)\beta}
\phi^I_{\alpha}{}^{\beta}
&&\nonumber\\
&\quad
+\left(
-\frac12G_{3IJK}
+2G_{4K(I,J)}
-12H_{1IJ(KLM)}X^{LM}
-8H_{1IJNLM,\langle OK\rangle}X^{LM}X^{NO}
\right)
&&\nonumber\\
&\hspace{20mm}\times
g_{\lambda(\mu}
\delta^{\lambda\alpha\beta}_{\nu)\gamma\delta}
\phi^I_\alpha\phi^{J\gamma}\phi^K_\beta{}^\delta
&&\nonumber\\
&\quad
-\frac14
\left(
G_4-2G_{4IJ}X^{IJ}
\right)
g_{\lambda(\mu}
\delta^{\lambda\alpha\beta}_{\nu)\gamma\delta}
R_{\alpha\beta}{}^{\gamma\delta}
&&\nonumber\\
&\quad
+\left(
\frac12G_{4IJ}
+X^{KL}G_{4IJKL}
\right)
g_{\lambda(\mu}
\delta^{\lambda\alpha\beta}_{\nu)\gamma\delta}
\phi^I_\alpha{}^\gamma\phi^J_\beta{}^\delta
&&\nonumber\\
&\quad
+\left(
\frac14G_{4IJ}
+2H_{2I(K|J|L)}X^{KL}
\right)
g_{\lambda(\mu}
\delta^{\lambda\alpha\beta\chi}_{\nu)\gamma\eta\delta}
\phi^I_\alpha\phi^{J\gamma}
R_{\beta\chi}{}^{\delta\eta}
&&\nonumber\\
&\quad
+\left[
\frac12G_{4IJKL}
+4H_{2I(K|J|L)}
+4H_{2IMJN,(KL)}X^{MN}
\right]
&&\nonumber\\
&\hspace{20mm}\times
g_{\lambda(\mu}
\delta^{\lambda\alpha\beta\chi}_{\nu)\gamma\eta\delta}
\phi^I_\alpha\phi^{J\gamma}
\phi^K_\beta{}^\delta\phi^{L\eta}{}_\chi
&&\nonumber\\
&\quad
+\frac18
\left(
\delta^{PQ}_{IK}\delta^{RS}_{JL}
+\delta^{PQ}_{JL}\delta^{RS}_{IK}
\right)
\Biggl[
\left(
H_{1PRQSN,\langle OM\rangle}
-2H_{1PRNSQ,\langle OM\rangle}
\right.
&&\nonumber\\
&\hspace{64mm}\left.
-2H_{1PNQSO,\langle RM\rangle}
\right)X^{NO}
&&\nonumber\\
&\hspace{42mm}
-\left(
H_{1PRQSM}
+H_{1PRMSQ}
+H_{1PRQMS}
\right)
\Biggr]
&&\nonumber\\
&\hspace{20mm}\times
g_{\lambda(\mu}
\delta^{\lambda\alpha\beta\chi}_{\nu)\gamma\eta\delta}
\phi^I_\alpha\phi^{J\gamma}
\phi^K_\beta\phi^{L\eta}\phi^M_\chi{}^\delta .
&&\tag{E.1}
\end{flalign}
As can be seen from this, we can confirm that the theory contains all basis functions up to the second order of the equations of motion for the bi-Galileon theory, as determined by Ohashi \textit{et al.} \cite{Ohashi:2015fma}.

\vspace{0.2cm}
\noindent


\let\doi\relax


\bibliographystyle{JHEP}
\bibliography{bib.bib}

\end{document}